\documentclass[10.5pt,letterpaper]{article}

\usepackage[T1]{fontenc}    
\usepackage[utf8]{inputenc} 
\usepackage{subfig,eurosym}
\usepackage[x11names]{xcolor}
\usepackage{amsmath,amssymb,amsfonts,amsthm}
\usepackage{mathtools,old-arrows,verbatim,mathrsfs,cite}
\usepackage{graphicx}
\usepackage{epstopdf,epsfig,changepage}

\usepackage{bm}
\usepackage[top=2.5cm, bottom=3cm, left=3.5cm, right=3.5cm,heightrounded,marginparwidth=1.5cm, marginparsep=1cm]{geometry}
\usepackage[shortlabels]{enumitem}
\usepackage{subcaption}
\usepackage[square,numbers,merge,comma,sort&compress]{natbib} 
\makeatletter
\def\NAT@spacechar{\,}  
\makeatother
\usepackage{comment}
\usepackage{relsize,setspace,moresize}
\usepackage{latexsym,mathrsfs,calligra,aurical}
\usepackage{float,appendix,xargs,extarrows,empheq,url}
\usepackage{physics}
\usepackage{cancel}

\usepackage{jheppub}
\hypersetup{
bookmarks=true,pdfmenubar=true,pdffitwindow=false,pdfpagemode={UseNone},pdfstartview={FitH},colorlinks=true,bookmarks=true,bookmarksnumbered=true,plainpages,linktoc=page,citecolor=blue,filecolor=black,
linkcolor=purple,urlcolor=blue}
\usepackage{footnotebackref}
\usepackage{hypernat}

\DeclareFixedFont\trfont{OT1}{phv}{b}{sc}{11}
\newcommandx{\overbar}[1]{\mkern1.5mu\overline{\mkern-2.0mu#1\mkern-2.0mu}\mkern1.5mu}

\DeclareFixedFont\trfont{OT1}{phv}{b}{sc}{11}
\DeclareMathAlphabet{\mathpzc}{OT1}{pzc}{m}{it}
\DeclareMathAlphabet{\mathcal}{OMS}{cmsy}{m}{n}
\DeclareSymbolFontAlphabet{\Scr}{rsfs}
\DeclareMathAlphabet{\mathbold}{U}{BOONDOX-ds}{m}{n}
\SetMathAlphabet{\mathbold}{bold}{U}{BOONDOX-ds}{b}{n}
\DeclareMathAlphabet{\mathcalboondox}{U}{BOONDOX-calo}{m}{n}
\SetMathAlphabet{\mathcalboondox}{bold}{U}{BOONDOX-calo}{b}{n}
\DeclareMathAlphabet{\mathbcalboondox}{U}{BOONDOX-calo}{b}{n}

\title{\centering\Large\bfseries{%
        Thermal superpotential and thermodynamics of neutral hairy black holes in extended SUGRA%
        }
\vspace{1.25em}}

\author[a,b]{Andrès Anabalòn,}
\emailAdd{andres.anabalon@uai.cl}

\author[c]{Dumitru Astefanesei,}
\emailAdd{dumitru.astefanesei@pucv.cl}

\author[d]{David Choque,}
\emailAdd{dchoque@pucp.edu.pe}

\author[e]{Antonio Gallerati}
\emailAdd{antonio.gallerati@polito.it}

\bigskip

\affiliation[a]{%
Universidad de Concepción, Departamento de Física, Casilla 160-C, Concepción, Chile
}

\affiliation[b]{%
Max Planck Institute for Gravitational Physics, Am Mühlenberg 1, D-14476 Potsdam, Germany
}

\affiliation[c]{Pontificia Universidad Cat\'{o}lica de Valpara\'{i}so,
Instituto de F\'{i}sica, Av.\ Brasil 2950, Valpara\'{i}so, Chile}

\affiliation[d]{Pontificia Universidad Católica del Perú. Av. Universitaria 1801, San Miguel, Perú}

\affiliation[e]{Politecnico di Torino, Dipartimento DISAT. Corso Duca degli Abruzzi 24, 10129 Torino, Italy}

\bigskip

\abstract{We present a family of exact neutral hairy black-hole solutions with spherical horizon topology in extended supergravity with Fayet--Iliopoulos terms. We consider a consistent dilaton truncation and analyze in detail a sector where the magnetic part of the FI terms vanishes. Using appropriate dilaton counterterms, we compute the thermodynamic quantities and show the existence of Hawking--Page phase transitions. As a holographic application, we derive the thermal superpotential in closed form and use it as a counterterm, explicitly demonstrating that no additional finite counterterms are required to regularize the Euclidean action and the quasi-local stress tensor. The dual stress tensor matches that of a thermal gas of massless particles and is consistent with mixed dilaton boundary conditions that preserve conformal symmetry.
}

\date{}

\begin{document}

\maketitle

\newpage

\section{Introduction}
Anti-de Sitter (AdS) black hole solutions are important in the context of the AdS--CFT duality. A well-understood limit of the duality is when, in the AdS bulk, it is sufficient to
consider the low-energy limit of the superstring theory, namely, supergravity (for a review, see \cite{Samtleben:2008pe,Gallerati:2016oyo}). 

Einstein-dilaton systems with a non-trivial dilaton potential provide an interesting class of gravitational theories that can be used to investigate the properties of dual field theories that are not supersymmetric and conformally invariant (see, e.g., \cite{Gursoy:2008za} and references therein). Since the radial coordinate in the bulk plays the role of energy in the dual field theory \cite{Peet:1998wn, Susskind:1998dq}, the gravitational equations should describe the holographic renormalization group (RG) flow \cite{Akhmedov:1998vf, Girardello:1998pd, Skenderis:1999mm, deBoer:1999tgo, Freedman:1999gp, Caceres:2025xzl}. This physical interpretation was extensively used to obtain a workable first-order formalism of the bulk gravity and a holographic $c$-function (see \cite{Aharony:1999ti} for an extensive review).\par\smallskip

In the following, we consider a consistent truncation of extended $\mathcal{N}=2$ gauged supergravity with abelian Fayet--Iliopoulos (FI) terms, retaining only the dilaton field with its self-interaction potential. This reduced model provides a well-defined subsector that admits exact neutral hairy black hole solutions \cite{Anabalon:2017yhv,Gallerati:2019mzs}, characterized by regular AdS asymptotics and controlled by the parameters of the dyonic gauging. Our analysis focuses on this particular class of solutions within the same supergravity framework. 

The chosen configurations are particularly interesting due to the mixed boundary
conditions satisfied by the dilaton, which are relevant in the dual field theory \cite{Witten:2001ua}. Deforming the gauge theory by adding relevant operators is one way to reduce its symmetries. Using the gravity side of the correspondence, one can obtain holographic RG flows corresponding to non-conformal field theories. The holographic interpretation of the hairy solutions with planar horizon geometry was presented in \cite{Anabalon:2020qux}, where the exact holographic RG flow was constructed. Concretely, the boundary perturbation generates a triple-trace deformation of the dual field theory. 

The purpose of this paper is to use the first-order formalism for hairy solutions that satisfy the equations of motion but are non-BPS, and to analyze an exact black hole family with spherical horizon geometry. We explicitly obtain the thermal superpotential and use it as a counterterm to regularize the Euclidean action and obtain the thermodynamic potential. Furthermore, a Hawking--Page transition \cite{Hawking:1982dh} exists at finite temperature consistently with the existence of the thermodynamic potential that corresponds to a confining theory \cite{Witten:1998qj}. The dual field theory lives on a spherical (boundary) geometry and the expectation value of its stress-energy tensor corresponds to a thermal gas of massless particles. 

We would like to point out an important difference between our present work and the planar hairy black hole family studied in \cite{Anabalon:2020qux}.
It is well known that, by using different foliations of AdS spacetime, it is possible to obtain boundaries that have different topologies affording the study of dual field theories living on different backgrounds. The diffeomorphisms
in the bulk are equivalent to the conformal transformations on the boundary and the
spherical boundary is related conformally to the flat boundary by a singular conformal transformation. This mathematical fact has an important consequence for the physical theory. While in our case there exist Hawking--Page phase transitions between the (large) hairy black hole with spherical horizon geometry and thermal AdS, there are no such transitions for the planar hairy black holes. However, for hairy black holes with toroidal horizon geometry, there exist first-order  phase transitions between the hairy black hole and the hairy AdS soliton \cite{Anabalon:2019tcy}.

The paper is structured as follows. In Section \ref{Sec2}, we describe the framework of a $\mathcal{N}=2$ supergravity model in the presence of an abelian FI dyonic gauge and obtain a consistent truncation to the dilaton. We present a general family of exact hairy black hole solutions for this SUGRA model. In Section \ref{Sec3}, we analyze exact hairy black holes in the pure electric FI sector and obtain their thermodynamic properties. In
Section \ref{thermalsup}, after a brief review of the first-order formalism and its generalization to black holes with spherical horizon geometry, we obtain an exact expression for the thermal superpotential and present some concrete applications within AdS--CFT duality. Finally, in Section \ref{Sec5}, we conclude with a discussion of our results.

\section{Gauged supergravity framework}
\label{Sec2}
The setup under analysis is an $\mathcal{N}=2$ supergravity model without hypermultiplets, containing a single vector multiplet whose scalar sector is parametrized by a complex field $z$. The theory is considered in the presence of an abelian FI dyonic gauging. Different instances of these models are distinguished by the geometry of the scalar manifold, which is determined by a prepotential depending on the complex scalar \cite{Gallerati:2016oyo,Gallerati:2019mzs}. Specifically, the special Kähler manifold is defined by the prepotential
\begin{equation} 
\mathcal{F}(\mathcal{X}^\Lambda)\,=-\frac{i}{4}\:\left(\mathcal{X}^0\right)^{n}\left(\mathcal{X}^1\right)^{2-n}\,, 
\label{eq:prepotentialn} 
\end{equation}
with $\mathcal{X}^\Lambda(z)$ denoting the components of a holomorphic section of the symplectic bundle over the scalar manifold. The projective coordinate $z$ is locally given by the ratio $\mathcal{X}^1/\mathcal{X}^0$ in the patch where $\mathcal{X}^0\neq 0$. 
Fixing $\mathcal{X}^0=1$, the holomorphic symplectic section $\Omega^M$ together with the Kähler potential take the form
\begin{equation} 
\Omega^M= \left(\begin{matrix} 1 \cr z \cr -\dfrac{i}{4}\,n\,z^{2-n} \cr -\dfrac{i}{4}\,(2-n)\,z^{1-n} \end{matrix}\right)\,, \quad\qquad 
e^{-\mathcal{K}}=\frac{1}{4}\,z^{1-n}\,\big(n\,z-(n-2)\,\bar{z}\big)\,+\,\text{c.c.} 
\end{equation}

The model is deformed by including abelian electric–magnetic FI terms encoded in a constant symplectic vector $\theta_M$:
\begin{equation} 
\theta_M=\left(\theta_1,\,\theta_2,\,\theta_3,\,\theta_4\right)\,, 
\label{eq:thetaM}
\end{equation}
which parameterizes the gauging. The scalar potential $V(z,\bar{z})$ then follows from
\begin{equation} 
V=\left(g^{i\bar{\jmath}}\,\mathcal{U}_i^M\,\overbar{\mathcal{U}}_{\bar{\jmath}}^N -3\,\mathcal{V}^M\,\overbar{\mathcal{V}}^N\right)\theta_M\,\theta_N
\,=\, -\frac{1}{2}\,\theta_M\,\mathcal{M}^{MN}\,\theta_N-4\,\mathcal{V}^M\,\overbar{\mathcal{V}}^N\theta_M\,\theta_N\;, 
\label{eq:Vpot} 
\end{equation}
where $\mathcal{V}^M=e^{\mathcal{K}/2},\Omega^M$, $\mathcal{U}_i^M=\mathcal{D}_i,\mathcal{V}^M$, and $\mathcal{M}$ is the symplectic, symmetric, negative-definite matrix encoding the non-minimal couplings of the scalar to the vector fields.

The complex scalar $z$ may be parametrized as
\begin{equation} 
z=e^{\lambda\,\varphi}+i\,\chi\,, 
\qquad\quad 
\lambda=\sqrt{\frac{2}{n(2-n)}}\;.\qquad 
\end{equation}

For our purposes, we focus on the truncation to the dilaton $\varphi$ only, imposing $\chi=0$. One can explicitly check that this truncation is consistent, as the axion equation of motion remains satisfied after setting $\chi=0$ \cite{Anabalon:2020pez}. To simplify the analysis, we perform the shift
\begin{equation} 
\varphi\;\rightarrow\;\varphi-\frac{2\,\nu}{\lambda\,(1+\nu)}\,\log(\theta_2\,\xi)\,, \label{eq:phishift} 
\end{equation}
which ensures that the potential has a minimum at $\varphi=0$ (cf.\ \eqref{eq:Vphi} and \eqref{eq:dVphi}). In this process, the FI parameters are redefined as
\begin{equation} 
\label{eq:thetanew} 
\theta_{1}=\frac{1+\nu}{-1+\nu}\;\theta_{2}^{-\frac{-1+\nu}{1+\nu}}\,\xi^{-\frac{2\,\nu}{1+\nu}}\,,
\qquad 
\theta_{3}=2\,\alpha\left(\xi\,\theta_{2}\right)^{\frac{-1+\nu}{1+\nu}}\,,
\qquad 
\theta_{4}=\frac{2\,\alpha}{\theta_{2}\,\xi}\,,
\end{equation}
with the auxiliary quantities
\begin{equation} 
\nu=\frac{1}{-1+n}\:,
\qquad\quad 
\xi=\frac{2\,L\,\nu}{-1+\nu}\,\frac{1}{\sqrt{1-\alpha^{2}\,L^{2}}}\:, 
\end{equation}
where $L$ denotes the AdS radius and $\alpha$ controls the strength of the dyonic gauging.

After applying the shift \eqref{eq:phishift}, the scalar field $z$ takes the form
\begin{equation} 
z=\left(\theta_{2}\,\xi\right)^{-\frac{2\,\nu}{1+\nu}}\,e^{\lambda\,\varphi}\,, 
\end{equation}
and the scalar potential \eqref{eq:Vpot} becomes explicitly \cite{Anabalon:2020pez}
\begingroup
\belowdisplayskip=-5pt
\belowdisplayshortskip=-5pt
\begin{equation} 
\begin{split} 
V(\varphi)\:=\,&-\frac{\alpha^2}{\nu^2}\,\left(\frac{(-1+\nu)(-2+\nu)}{2}\, e^{-\varphi\,\ell\,(1+\nu)} + 2\,(-1+\nu^2)\,e^{-\varphi\,\ell} +\frac{(1+\nu)(2+\nu)}{2}\,e^{\varphi\,\ell\,(-1+\nu)}\right)+ \\ &+\frac{\alpha^2-L^{-2}}{\nu^2}\,\left(\frac{(-1+\nu)(-2+\nu)}{2}\, e^{\varphi\,\ell\,(1+\nu)} + 2\,(-1+\nu^2)\,e^{\varphi\,\ell} +\frac{(1+\nu)(2+\nu)}{2}\,e^{-\varphi\,\ell\,(-1+\nu)}\right), \end{split} \label{eq:Vphi} 
\end{equation}
\endgroup
where $\ell=\tfrac{\lambda}{\nu}=\tfrac{1}{\nu}\,\sqrt{\tfrac{2\,\nu^2}{-1+\nu^2}}$, and the parameter $\theta_2$ has been eliminated via the above redefinitions. 
%
%

\par\smallskip

The $\nu$ real parameter, $\abs{\nu}>1$, interpolates among all single-dilaton truncations of maximal $\mathrm{SO}(8)$ supergravity in four dimensions \cite{deWit:1981sst,deWit:1982bul,DallAgata:2012mfj}. Depending on its value, the $\mathrm{SO}(8)$ gauge symmetry is broken according to \cite{Anabalon:2020pez,Gallerati:2021cty}: 
\begin{equation}
\label{eq:SO8breaking} 
\begin{alignedat}{3} 
\nu=-2 & \;\;\rightarrow\quad && \mathrm{SO}(6)\times\mathrm{SO}(2) \,, 
\\[1ex]
\nu=4 & \;\;\rightarrow\quad && \mathrm{SO}(5)\times\mathrm{SO}(3) \,, 
\\[1ex] 
\nu=\infty & \;\;\rightarrow\quad && \mathrm{SO}(4)\times\mathrm{SO}(4) \,, 
\\[1ex] 
\nu=4/3 & \;\;\rightarrow\quad && \mathrm{SO}(7) \,,
\end{alignedat} 
\end{equation} 
while for generic $\abs{\nu}>1$ the resulting models correspond to $\mathcal{N}=2$ supergravity theories.
%

%
%

\paragraph{Electric gauging.}
In the following section \ref{subsec:hairyBHsol}, we will consider a hairy black hole configuration in the described theory, corresponding to Family 2 of hairy solutions of \cite{Anabalon:2017yhv}. 
We will focus on solutions with spherical horizon topology that admit a well-defined electric gauging limit $\alpha=0$, corresponding to the vanishing of the magnetic part of the FI terms in \eqref{eq:thetaM}, see \eqref{eq:thetanew}. 

In general, the truncation to the dilaton $\chi=0$ is consistent at the level of the scalar potential $V$, but not of the superpotential. In fact, if we consider the real superpotential $\mathcalboondox{W}=\lvert\mathcal{W}\rvert$, one generically finds \,$\partial_\chi\mathcalboondox{W}\big|_{\chi=0}\!\neq0$\,; therefore, the dilaton truncation is not consistent at the level of $\mathcalboondox{W}$ since the potential could not be expressed in terms of \,$\mathcalboondox{W}\big|_{\chi=0}$ and its derivatives with respect to $\varphi$. However, in the purely electric gauging $\alpha=0$, the truncation remains consistent also at the level of the real superpotential, since \,$\partial_\chi \mathcalboondox{W}\big|_{\chi=0}\!=0$\, in this limit \cite{Anabalon:2020qux}.


\subsection{The model}
The model under consideration is an Einstein-dilaton model endowed with a particular dilaton potential. It is described by the action
\begin{equation}
S=-\frac{1}{8\pi G} \,\int_{\mathcal{M}}\!d^{4}x\,\sqrt{-g}\,\left(\frac{R}{2}-\frac{1}{2}
\left(\partial\varphi\right)^2+V(\varphi)\right)\,,
\end{equation}
with the dilaton potential \eqref{eq:Vphi}. 
We are going to use the same notation as in \cite{Anabalon:2017yhv} and so the cosmological constant $\Lambda$ is related to the AdS radius $L$ as usual $(\Lambda=- 3/L^{2})$.

The scalar potential \eqref{eq:Vphi} is invariant under the change
\begin{equation} \label{eq:ST}%
\varphi\,\rightarrow\,-\varphi\,,\qquad
\alpha^{2}\,\rightarrow\,L^{-2}-\alpha^{2}\,,
\end{equation} 
and, as mentioned, two families of solutions turn out to be related by this non-perturbative electric–magnetic duality -- a global symmetry of the ungauged theory that extends to the gauged case when the constant theta term transforms accordingly; the dual configurations satisfy the same field equations for dual values of $\alpha$, as shown in \cite{Anabalon:2020pez}.%
\footnote{%
Since the symmetry acts non-trivially on the FI parameters, it provides a (non-universal) effective solution-generating mechanism \cite{Breitenlohner:1987dg,Cvetic:1995kv,Gaiotto:2007ag,Bergshoeff:2008be,Bossard:2009at,Andrianopoli:2013kya,Andrianopoli:2013jra} in asymptotically AdS spacetimes, where the duality action on the embedding tensor otherwise obstructs the use of global isometries \cite{Gallerati:2016oyo,Gallerati:2019mzs}.}
The theory features a vacuum at
$\varphi=0$, satisfying
\begin{equation}
	V(0)=-\frac{3}{L^{2}}\;,
    \qquad\quad
    \left.\frac{dV(\varphi)}{d\varphi}\right\vert
	_{\varphi=0}=0\;,
    \qquad\quad
    \left.\frac{d^{2}V(\varphi)}{d\varphi^{2}}\right\vert
	_{\varphi=0}=-\frac{2}{L^{2}}\;.
\label{eq:dVphi}
\end{equation}

Note that the mass of the scalar field is $m^{2}=-2/L^{2}$. The model is invariant under $\nu\leftrightarrow-\nu$ and so we can restrict the analysis to $1\leq\nu<\infty$. The equations of motion for the metric and dilaton are
\begin{equation}
\begin{split}
&E_{\mu\nu}=G_{\mu\nu}-\frac{1}{2}T_{\mu\nu}^{\varphi}=0\,, \qquad
T_{\mu\nu}^{\varphi}=\partial_{\mu}\varphi\,\partial_{\nu}\varphi-g_{\mu\nu}\left(\frac{1}{2}(\partial\varphi)^{2}-V(\varphi)\right),
\\[1.5ex]
&\frac{1}{\sqrt{-g}}\,\partial_{\mu}\left(\sqrt{-g}\,g^{\mu\nu}\partial_{\nu}\varphi\right)-\frac{dV}{d\varphi}=0\:.
\end{split}
\end{equation}
\vspace{\parskip}

\subsection{The general hairy black hole solution}
\label{subsec:hairyBHsol}
In what follows, we focus on Family $2$ of hairy solutions presented in \cite{Anabalon:2017yhv}.%
\footnote{In the presence of electric field, there exist asymptotically flat hairy black 
holes in the same family (with the same dilaton potential), which are thermodynamically and dynamically 
stable \cite{Anabalon:2013qua, Astefanesei:2019mds, Astefanesei:2019qsg, Astefanesei:2020xvn} and, also, have non-trivial critical behavior \cite{Astefanesei:2025tpn}. 
} 
This new class of exact hairy black hole solutions was obtained by applying
the duality transformation corresponding to the supergravity potential directly to the first family of hairy solutions. Although some basic properties of these new solutions were only briefly mentioned in \cite{Anabalon:2017yhv}, we shall analyze them in detail in the present work. The main reason we are interested in the second Family is that the solutions are still regular in the limit $\alpha=0$, which is consistent with the existence of a real superpotential in the extended supergravity model discussed
in the previous section.

The general ansatz for the metric, for both families, is
\begin{equation} \label{eq:AdSmetr}
ds^{2}=
 \Upsilon(x)\,\left(f(x)\,dt^{2}-\frac{\eta^2}{f(x)}\,dx^2-d\Sigma_{k}\right)\;,
\end{equation}
\sloppy
where $x$ is the radial variable and $d\Sigma_{k}$ is the metric on the horizon. The latter is a two-dimensional manifold with constant curvature, fully characterized by its Ricci scalar  \,${R_{k}=2 k/L^2}$, with \,${k=0,-L^{-2},+L^{-2}}$\,. The parameter $\eta$ is the unique integration constant of the system non-trivially related to the mass. The engineering dimension of $\eta$ is $\left[\eta\right]=\left[M\right]^{-1}=\left[L\right]$. %
Without loss of generality, we are going to consider from now on a positive definite  $\eta>0$.\par
Using the integration techniques that are described in \cite{Anabalon:2017yhv}, one obtains a general class of regular exact hairy 
black hole solutions characterized by the functions $\varphi(x),\,\Upsilon(x),\,f(x)$ as
\begin{equation} \label{eq:AdSfunc}
\begin{split}
\varphi(x)&=-\ell^{-1}\,\ln(x)\;, \qquad\quad\ell=\frac{1}{\nu}\,\sqrt{\frac{2\,\nu^2}{-1+\nu^2}}\;, \qquad\quad\;
\Upsilon(x)=\frac{L^2\nu^2\,x^{\nu-1}}{\eta^2\,(x^\nu-1)^2}\;, 
\\[3.5\jot]
f(x)&=\frac{x^{2-\nu}\,(x^\nu-1)^2\,\eta^2\,k}{\nu^2} +
1+\alpha^{2}L^2\,\left(-1+\frac{x^2}{\nu^2}\,\left((\nu+2)\,x^{-\nu}-(\nu-2)\,x^\nu+\nu^2-4\right)\right).
\end{split}
\end{equation}

However, solutions with a spherical horizon topology, $k=L^{-2}$,  are not regular when the magnetic gauging vanishes
($\alpha=0$) and so become naked singularities in this limit.  \par\smallskip

The focus of our work is on a different class of which we will describe in the following. Being that the theory is invariant under the change \eqref{eq:ST}, we perform the transformation on the first family \eqref{eq:AdSfunc} to find the following configuration:
\begin{equation}
\begin{split}
\varphi&=\ell^{-1}\,\ln(x)\;, \qquad\quad\;  \ell=\frac{1}{\nu}\,\sqrt{\frac{2\,\nu^2}{-1+\nu^2}} \;, \qquad\quad\;
\Upsilon(x)=\frac{L^2 \nu^2\,x^{\nu-1}}{\eta^2\,(x^\nu-1)^2}\;, 
\\[3.5\jot]
f(x)&=\frac{x^{2-\nu}\,\left(x^\nu-1\right)^2\,\eta^2\,k}{\nu^2} +
      1+\left(1-\alpha^2 L^2\right)\,\left(-1+\frac{x^2}{\nu^2}\,\left((\nu+2)\,x^{-\nu}-(\nu-2)\,x^\nu+\nu^2-4\right)\right).
\end{split}
\end{equation}

The asymptotic region of the theory is located at the pole of order 2 of the conformal factor, namely $x=1$. The geometry and scalar field are singular in $x=0$ and $x=\infty$. Therefore, the configuration contains two disjoint geometries given by $x\in(1,\infty)$ or $x\in(0,1)$. These geometries have an invariant characterization in terms of the scalar field:
\begin{equation}
\begin{split}
x\in(0,1) \;\;&\Longrightarrow\quad \varphi\leq0\;, \\
x\in(1,\infty) \;\;&\Longrightarrow\quad \varphi\geq0\;,
\end{split}
\end{equation}
and we shall call them the negative branch and the positive branch accordingly.
Interestingly, in the limit $\alpha=0$, the black hole solutions with spherical horizon topology of Family 2 remain regular with a well-defined horizon only in the positive branch.

\section{Thermodynamics of hairy black holes in pure electric FI sector}
\label{Sec3}
In this section, we analyze in great detail the exact hairy black hole solutions when $\alpha=0$. We focus on the hairy black holes in the positive branch, with $k=1/L^2$, for which we explicitly prove the existence of a regular horizon. We find that the mixed boundary conditions for the dilaton are compatible with AdS asymptotics, and thus the conformal symmetry is preserved. Then, we use dilaton counterterms \cite{Marolf:2006nd,Anabalon:2015xvl} to regularize the Euclidean action, obtain the thermodynamic quantities, and verify the first law of black hole thermodynamics and quantum statistical relation. In the last part of this section we use the free energy to prove the existence of first-order phase transitions.

\subsection{Hairy black hole and its properties}
For clarity of exposition, let us now collect together all the relevant results of the previous section, but for the particular
case we are interested in, namely $\alpha=0$ and $k=1/L^2$. Specifically, in the positive branch $x>1$, the solution becomes
\begin{equation}
\begin{split}
\varphi&=\ell^{-1}\,\ln(x)\;, \qquad\quad\;  \ell=\frac{1}{\nu}\,\sqrt{\frac{2\,\nu^2}{-1+\nu^2}} \;, \qquad\quad\;
\Upsilon(x)=\frac{L^2 \nu^2\,x^{\nu-1}}{\eta^2\,(x^\nu-1)^2}\;,\qquad
\\[\jot]
f(x)&=\frac{x^{2}}{\nu^{2}}\left((\nu+2)\,x^{-\nu}-(\nu-2)\,x^{\nu}+\nu^{2}-4\right)+\frac{x}{\Upsilon(x)}\:,
\end{split}
\end{equation}
and the dilaton potential, drastically simplified, is
\begin{equation}
\label{potential1}
	V(\varphi)=-
\frac{1}{L^{2}\nu^{2}} \left(  \frac{\left(
	\nu-1\right)  \left(  \nu-2\right)  }{2}\,e^{\varphi \,\ell\,\left(
	\nu+1\right)  }+2\left(  \nu^{2}-1\right)  \,e^{\varphi\,\ell}+\frac{\left(
	\nu+1\right)  \left(  \nu+2\right)  }{2}\,e^{-\varphi\, \ell\,\left(\nu-1\right)  }\right)
\end{equation}

We would like to emphasize that $L$ is the radius of AdS and $\ell$ is an expression depending of the `hair' parameter $\nu$ we use to simplify the equations.

The location of the horizon, $x_\text{h}$, can be obtained when $f(x_\text{h})=0$, which can be solved analytically, and we get the following solutions:
\begin{equation}
2\left(x_\text{h}\right)^\nu = \frac{2\,\eta^{2}L^{-2} - (\nu^{2}-4)\pm\nu\sqrt{\nu^{2}-4-4\,\eta^{2}L^{-2}}}{\eta^{2}L^{-2}-(\nu-2)}\:. 
\end{equation}

The horizon is present only when $\nu > 2$, for which the solution with the $+$ sign is the only one that is positive definite.\par\smallskip

Now that we have proved that the solution is regular, let us obtain the corresponding boundary conditions. While the $x$-coordinate is useful for integrating the equations of motion and obtaining the hairy black hole solution in analytic form, one has to use the canonical coordinates (namely, the usual radial coordinate) to obtain some physical properties. Therefore, the change of coordinates we are interested in corresponds to
\begin{equation}
	\Upsilon(x)=\frac{r^{2}}{L^{2}}+O\left(r^{-2}\right) \quad \Longrightarrow \quad x=1 +\left(\frac{L^2}{r \eta}+L^6\,\frac{1-\nu^2}{24\,(r \eta)^3}\right)+L^8\, \frac{\nu^2-1}{24\,(r \eta)^4}+...
\end{equation}

It is now straightforward to obtain the asymptotic expansion in the canonical coordinates for the dilaton  as
\begin{equation}  \label{scalarfalloff}
\varphi=L^2\,\frac{\varphi_0}{r}+L^4\,\frac{\varphi_1}{r^2}%
+O\left(r^{-3}\right)= L^2\,\frac{1}{\ell\,\eta\,r}-L^4\,\frac{1}{%
2\,\ell\,\eta^2\,r^2}+ O\left(r^{-3}\right)\;,
\end{equation}
with
\begin{equation}
\label{bconditions}
	\varphi_{0}=\frac{1}{\ell\eta}\,, \quad
	\varphi_{1}=-\frac{1}{2\ell\eta^{2}} \quad \Longrightarrow 	\quad \varphi_{1}=-\frac{\ell\,\varphi_{0}^{2}}{2}\,.
\end{equation}

Similarly, when expressed in the canonical coordinates, the form of the metric takes the following form
\begin{equation}
ds^{2}=g_{tt}(r)\,dt^{2}+g_{rr}(r)\,dr^{2}-\left(\frac{r^{2}}{L^{2}}+O\left(r^{-2}\right)\right)d\Sigma_{1}\:,
\end{equation}
with the following fall-off for its components: 
\begin{align}
\label{metricfalloff}
	g_{tt}  & =\frac{r^{2}}{L^{2}} + 1 - \frac{\mu L^{4}}{r}+O\left(r^{-2}\right)\:,
    \\[\jot]
	-g_{rr}  & =\frac{L^{2}}{r^{2}}+\frac{L^6}{r^{4}}\left(-\frac{1}{L^2}-\frac{\varphi_{0}^{2}}{2}\right)
	+\frac{L^{9}}{r^{5}}\left(\frac{\mu}{L}-\frac{4\,\varphi_{0}\,\varphi_{1}}{3L}\right)+O\left(r^{-6}\right)\:.
\end{align}

Since the mixed boundary conditions for the dilaton preserve the conformal symmetry on the boundary, 
we define the mass parameter
\begin{equation} 
\mu \equiv -\frac{1}{\eta L^2}+\frac{\nu^{2}-4}{3\eta^{3}}
\end{equation}
such that the total mass, as can be read from the $g_{tt}$ component of the metric, with the right normalization, is
\begin{equation}
M=\frac{\sigma_{1}L^{4}\mu}{8\pi G}\:,
\end{equation}
where $\sigma_{k}=L^{-2}\int{\!d\Sigma_{k}}$ (in our case, $k=L^{-2}$, \,$\sigma_{+1}=4\pi$).
As a consistency check, we observe that the total mass is positive defined if and only if~$\nu>2$,
\begin{equation}
\label{mass}
M=\frac{L^{4}}{2 G}\left(-\frac{1}{L^{2}\eta}+\frac{\nu^{2}-4}{3\eta^{3}}\right)>0 \quad\Longrightarrow \quad 
 \nu>2\,, ~~ \text{with}~~ k=L^{-2}\:,
\end{equation}
which is the same condition that ensures the regularity of the hairy black hole solution. The same result will be obtained in the next section using the counterterm method and regularized quasi-local stress tensor of Brown and York \cite{Brown:1992br}.

\sloppy
The regularity condition can be obtained from the horizon equation $f(x_\text{h},\alpha,\eta,\nu)=0$, with ${\nu=2+\epsilon^{2}}$ and large horizons:
\begin{equation}
\underbrace{\frac{1}{x_\text{h}^{\epsilon^2}}\left(\frac{\eta^{2}L^{-2}+\nu+2}{\nu^{2}}\right)}_{x_\text{h}\rightarrow\,\infty}
+\underbrace{x_\text{h}^{2}\left(\frac{-2\eta^{2}L^{-2}+\nu^{2}-4}{\nu^{2}}\right)+x_\text{h}^{\epsilon^{2}+4}\left(\frac{\eta^{2}L^{-2}-\nu+2}{L^{2}\nu^{2}}\right)}_{x_\text{h}^{2}(\ldots)\,\ll\: x_\text{h}^{\epsilon^{2}+4}(\ldots)} =0\:. 
\end{equation}

Then, we obtain a critical value for the integration constant that is well defined only for the hair parameter interval $\nu>2$:
\begin{equation} x_\text{h}^{\epsilon^{2}+4}\left(\frac{\eta^{2}L^{-2}-\nu+2}{L^{2}\nu^{2}}\right)=0 \quad\Longrightarrow \quad
\eta_\text{cr}=L\,\sqrt{\nu-2}\:.
\label{F2}
\end{equation}

We would like to mention that the holographic interpretation of dilaton's mixed boundary conditions and hairy black hole solution in the context of AdS--CFT duality will be provided in subsequent Section \ref{thermalsup}.

\subsection{Dilaton counterterms and thermodynamics}
\label{DilatonCounterterm}
The counterterm method in AdS was developed in \cite{Skenderis:2000in,Henningson:1998gx,Balasubramanian:1999re,Mann:1999pc,Emparan:1999pm}. We are going to closely follow \cite{Marolf:2006nd,Anabalon:2015xvl}, where specific counterterms for the scalar fields were used to regularize the action and obtain the holographic mass. 
Although in theories of gravity coupled to matter the theory is usually fully determined
by the action, there is an important subtlety in the presence of scalar fields with tachyonic mass in AdS. That is, when both modes in the scalar field’s fall off  (\ref{scalarfalloff}) 
are normalizable \cite{Balasubramanian:1998sn}, they represent physically acceptable fluctuations. Therefore, specifying boundary conditions for the scalar
field is equivalent to fixing the boundary data $(\varphi_{0}, \varphi_{1})$ or a specific relation between them. It is
common to denote the mixed boundary conditions on the scalar field by $\varphi_{1}=d{w}/d{\varphi_{0}}$, where $w({\varphi_{0}})$ is an arbitrary differentiable function. This constraint on $\varphi_{0}$ and $\varphi_{1}$ can be obtained
from the vanishing symplectic flux flow through the boundary \cite{Amsel:2006uf} and is interpreted as an integrability condition for the mass in the Hamiltonian formalism \cite{Hertog:2004ns,Anabalon:2014fla}.

In this subsection, we are going to compute the on-shell Euclidean action and show that the quantum statistical relation is satisfied. We consider the general ansatz
\begin{equation}
\begin{split}
ds^{2}=N(r)\,dt^{2}-H(r)\,dr^{2}-S(r)\,d\Sigma_{1}\:,
\qquad\quad
d\Sigma_{1}=d\theta^{2}+\sin^{2}\theta\,d\phi^{2}\:,
\end{split}
\end{equation}
with the fall-off (\ref{metricfalloff}). With this metric ansatz, we are going to evaluate the full
regularized action that consists of the bulk part, Gibbons-Hawking boundary term $I_{GH}$,
gravitational counterterm for the asymptotically AdS spacetime $I_g$, and a boundary term
for the scalar field $I_{\varphi}$, namely
\begin{equation}
\label{action}
I^{\textsc{e}}\left(g^{\textsc{e}},\varphi\right)=
\frac{1}{8\pi G}\int_{M}d^{4}x\,\sqrt{g^{\textsc{e}}}\,\left( \frac{R}{2}-\frac{1}{2}\,\left(\partial\varphi\right)^{2}+V(\varphi)\right) - \frac{1}{8\pi G}\,\int_{\partial M}d^{3}x\,\sqrt{h}\,K + I_\mathrm{g}^{\textsc{e}} + I_{\varphi}^{\textsc{e}}\:,
\end{equation}
where
\begin{equation}
\label{Igrav}
\begin{split}
I_\mathrm{g}^{\textsc{e}}=\frac{1}{8\pi G}\,\int_{\partial M} d^{3}x\,\sqrt{h}\,\left(
\frac{2}{L}+\frac{L}{2}R(h)\right),
\end{split}
\end{equation}
and 
\begin{equation}
I_{\varphi}^{\textsc{e}}=\frac{1}{8\pi G}\int_{\partial M}d^{3}x\,
\sqrt{h}\left(\frac{\varphi^{2}}{2L}+\frac{w(\varphi_{0})}{L\varphi_{0}^{3}}\varphi^{3}\right)=\frac{1}{8\pi G}\int_{\partial M}d^{3}x\,
\sqrt{h}\left(\frac{\varphi^{2}}{2L}-\frac{\ell \varphi^{3}}{6L}\right).
\end{equation}

Here, we denote the induced metric on the slice $r=\text{constant}$ by $h_{ab}$, the Ricci scalar of the $3$-dimensional surface by $R(h)$, and the trace of the extrinsic curvature as $K$. 

To facilitate the comparison with the results of Section \ref{thermalsup}, we split the computation of the Euclidean 
action into two parts. We evaluate the first three terms and identify the divergent terms,
\begin{equation}
I^{\textsc{e}}_\text{bulk}+I^{\textsc{e}}_{\textsc{gh}}+I_\text{g}^{\textsc{e}}=-\frac{\mathcal{A}}{4G}+\frac{L^{2}}{2 GT}\left(L^{2}\mu+\frac{2\ell L^{2}}{3}\varphi_{0}^{3}-\frac{r\varphi_{0}^{2}}{2}\right)
\end{equation}
and, separately, the contribution from the counterterm associated to the dilaton,
\begin{equation}
I_{\varphi}^{\textsc{e}}=\frac{L^{2}}{2 GT}\,N^{1/2}\,S\left(\frac{\varphi^{2}}{2L}-\frac{\ell \varphi^{3}}{6L}\right)=\frac{L^{2}}{2 GT}\left(-\frac{2\ell L^{2}}{3}\varphi_{0}^{3}+\frac{r\varphi_{0}^{2}}{2}\right)
\end{equation}
where $T$ is the temperature and area of the event horizon is denoted as $\mathcal{A}$. 

Clearly, the divergent terms cancel each other in the final expression of the total Euclidean action and we 
obtain
\begin{equation}
\label{actionF}
F=I^{\textsc{e}}T=
\frac{ L^{4}}{2 G}\left(-\frac{1}{L^2 \eta}+\frac{\nu^{2}-4}{3\eta^{3}} \right) - \frac{T\mathcal{A}}{4G}\:.
\end{equation}

As expected, the first term matches the mass (\ref{mass}), though, for completeness, we will get the mass using the quasi-local stress tensor obtained from the variation of the total action, including the counterterms, in the Lorentzian section:
\begin{equation}
\label{BY}
\tau_{a b}=\frac{1}{8 \pi G}\left(K_{a b}-h_{a b} K+\frac{2}{L} h_{a b}-L G_{a b}\right)+\frac{h_{a b}}{8 \pi G}\left(\frac{\varphi^2}{2 L}-\frac{\ell \varphi^3}{6 L}\right)\,.
\end{equation}

Once the quasi-local stress tensor is known, the conserved quantities can be obtained, provided the quasi-local surface has an isometry generated by a Killing vector. The relevant component is
\begin{equation}
\tau_{tt}=-\frac{1}{8\pi G}\left[\frac{NS^{\prime}}{2S\sqrt{H}}-\frac{2N}{L}\left(1+\frac{1}{2S}\right)\right]+\frac{N}{8\pi G}\left(\frac{\varphi^{2}}{2L}-\frac{\ell \varphi^{3}}{6L}\right)\,,
\end{equation}
and the energy of the gravitational system, which is the conserved charge associated with the Killing vector $\xi^{b}=\delta^{b}_{t}$ due to the time translational symmetry of the metric tensor, is
\begin{equation}
E=\oint_{s^2_\infty}d^{2}y\,\sqrt{\sigma}\,n^{a}\,
\tau_{ab}\,\xi^{b}=\left(\oint d\theta\,d\phi\,\sin{\theta}\right)\frac{SL^{2}}{\sqrt{N}}\tau_{tt} = \frac{L^{4}\mu}{2 G}\:,
\end{equation}
where $n^{a}=\delta^{a}_{t}/\sqrt{g_{tt}}$ is the unit normal to the surface $t = \text{constant}$, $s^2_\infty$, at spatial infinity, and $\sigma$ is the determinant of the induced metric on $s^2_\infty$.

Finally, eq.\ (\ref{actionF}) provides the correct quantum statistical relation, where $x_\text{h}$ is the horizon localization $f(x_\text{h})=0$
\begin{equation}
\label{F}
	F=M-TS=\frac{L^{4}}{4\pi G}\left(\frac{\nu\left(x_\text{h}^{\nu}+1\right)}{\left(x_\text{h}^{\nu}-1\right)\eta \,L^2}-\frac{\nu^{2}-4}{3\eta^{3}}\right)
\end{equation}
once we identify the free energy (\ref{actionF}) as $F=I^{\textsc{e}}T$ and the temperature, $T$, as the inverse of the periodicity of the Euclidean time after removing the conical singularity,
\begin{equation}
	T=\frac{L^{2}}{4\pi\eta\Upsilon}\left(-{\frac { {x}^{\nu}\nu+2\,x_\text{h}^{\nu}+\nu-2 }{L^2\left(x_\text{h}^{\nu}-1\right)
	}}+{\frac {{\nu}^{2}-4}{{\eta}^{2}
	}}
	\right)\:.
\end{equation}

\subsection{Hawking--Page phase transitions}
\begin{figure}[!ht]
\centering
\subfloat{
\includegraphics[width=0.45\textwidth]{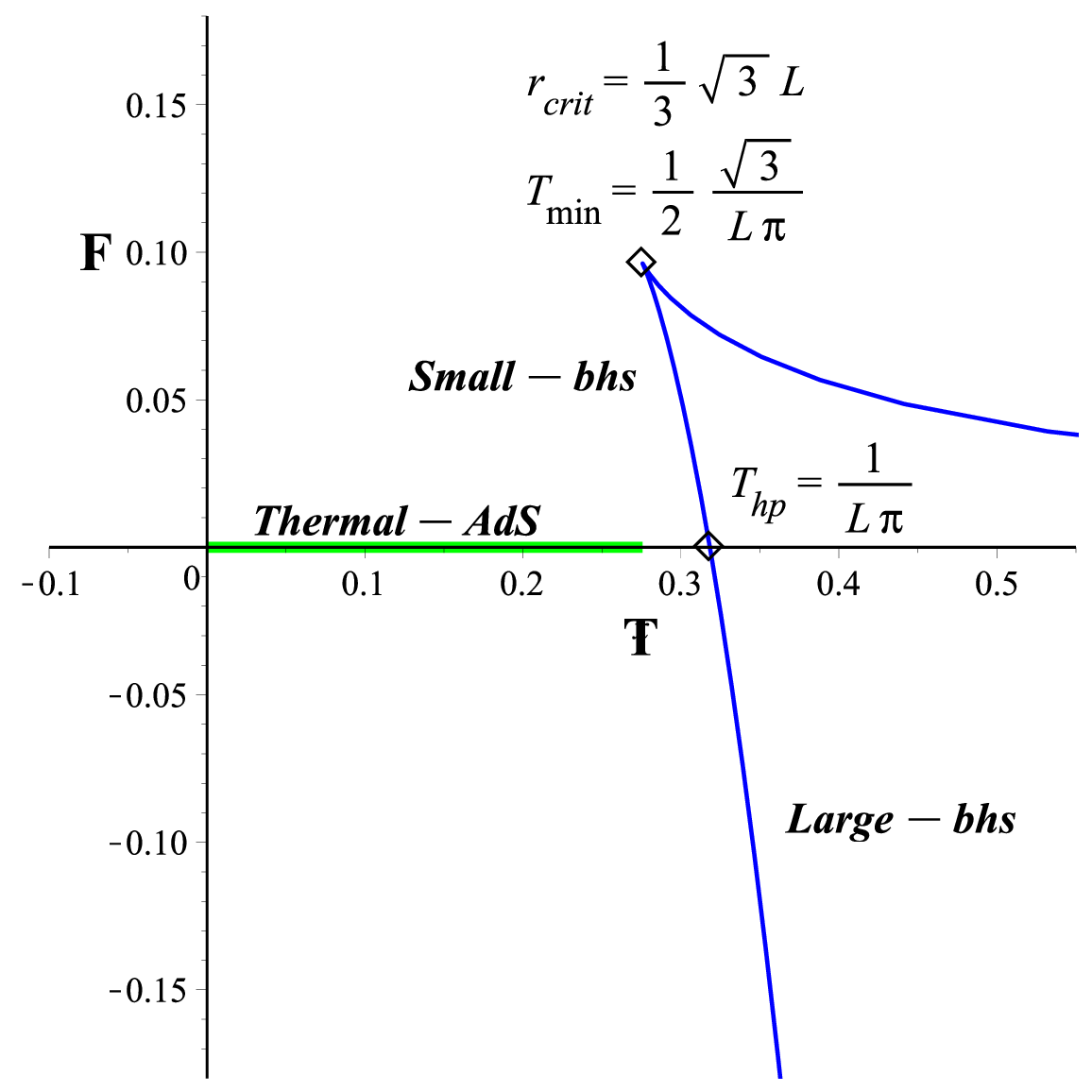}}
\hfill
\subfloat{
\includegraphics[width=0.45\textwidth]{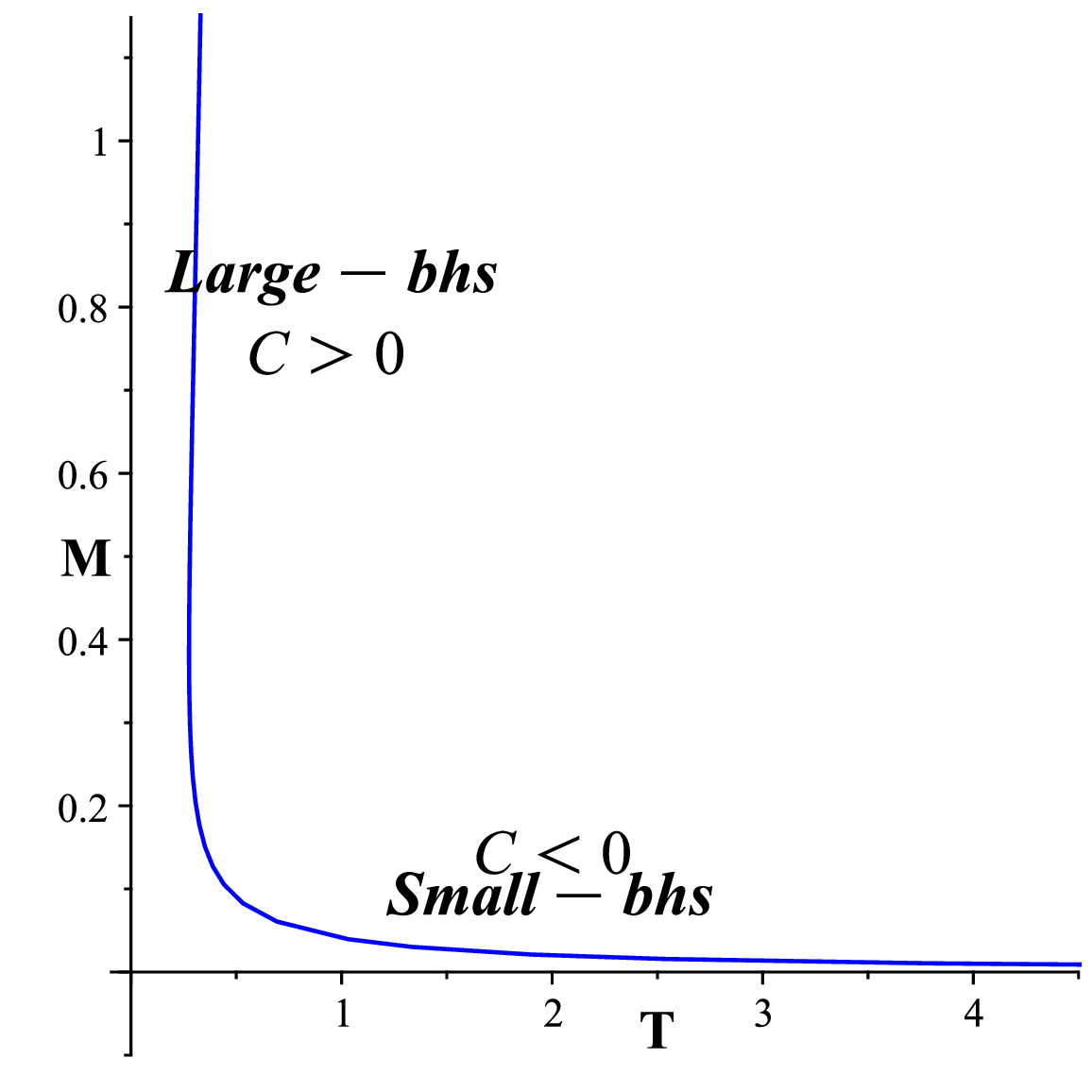}}
\caption{The left panel shows the free energy as a function of temperature, exhibiting a first-order phase transition at $F = 0$, corresponding to the Hawking--Page temperature $T_\textsc{hp} = 1/(\pi L)$. The right panel displays the mass-temperature relation, which features two branches corresponding to small (unstable) and large (stable) black holes.
} 	
\label{fig1}  
\end{figure}

Let us first consider the limit $\nu\to1$. Although sending $\varphi \rightarrow 0$ is not consistent with the equations of motion, the branch reduces to Schwarzschild–AdS (up to a parameter redefinition) when the hair parameter is set to unity. In this case, the dilaton potential (\ref{potential1}) matches the usual cosmological constant of AdS$_4$. The thermodynamics of Schwarzschild--AdS black hole was studied by Hawking and Page in \cite{Hawking:1982dh}, where they proved that the large black holes in AdS are thermodynamically stable and there are phase transitions of first-order between thermal AdS and the large black hole. These results are summarized in Fig.~\ref{fig1}.

We shall proceed with a similar analysis for hairy black holes, keeping in mind that the results are not analytic, and we are going to use parametric equations to obtain the corresponding plots. Again, we are going to focus on the case $\alpha=0$, $\varphi>0$ and $1\leq x<\infty$. Let us first rescale the physical quantities to obtain the following dimensionless expressions:
\begin{equation}
\left(\frac{MG}{L},~\frac{FG}{L},~\frac{SG}{L^{2}\pi},~TL\pi\right) \;\;\Rightarrow\;\; \left(M,~F,~S,~T\right)
\end{equation}

The horizon equation can be used to solve for $\eta$ (with $x_\text{h}>1$, $\nu>2$) for which we get
\begin{equation}
f(x_\text{h},\eta)=0 \quad\Rightarrow\quad \eta(x_\text{h},\nu) = \frac{\sqrt{x_\text{h}^{2\nu} (\nu - 2) - x_\text{h}^\nu (\nu^2 - 4) - (\nu + 2)}}{x_\text{h}^\nu - 1}\:,
\end{equation}
such that the dimensionless thermodynamic quantities become
\begin{align}
M(x_\text{h},\nu)&= \frac{1}{2} \left( -\frac{1}{\eta} + \frac{\nu^2 - 4}{3\eta^3} \right), \qquad S(x_\text{h},\nu) = \frac{\nu^2 x_\text{h}^{\nu - 1}}{\eta^2 (x_\text{h}^\nu - 1)^2}\:,
\\[1.5ex]
T(x_\text{h},\nu)&= -\frac{\eta}{4} \left(\frac{\nu^2 x_\text{h}^{\nu -1}}{(x_\text{h}^\nu - 1)^2} \right)^{-1} 
\left( \frac{x_\text{h}^\nu \nu + 2x_\text{h}^\nu + \nu - 2}{x_\text{h}^\nu -1} - \frac{\nu^2 - 4}{\eta^2} \right),
\\[2ex]
F(x_\text{h},\nu) &= \frac{1}{4\eta} \left(\nu \, \frac{x_\text{h}^\nu + 1}{x_\text{h}^\nu - 1} - \frac{\nu^2 - 4}{3\eta^2} \right).
\end{align}

\begin{figure}[!ht]
\centering
\subfloat{
\includegraphics[width=0.45\textwidth]{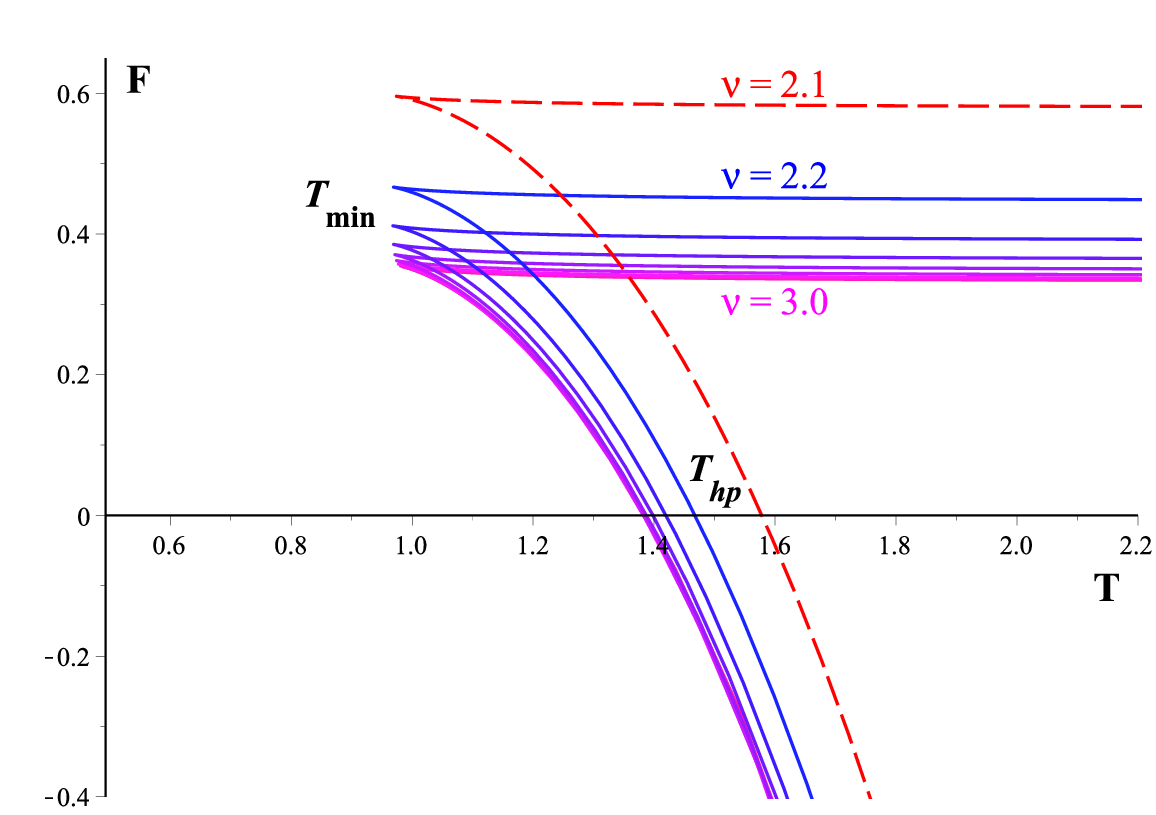}
}
\hfill
\subfloat{
\includegraphics[width=0.45\textwidth]
{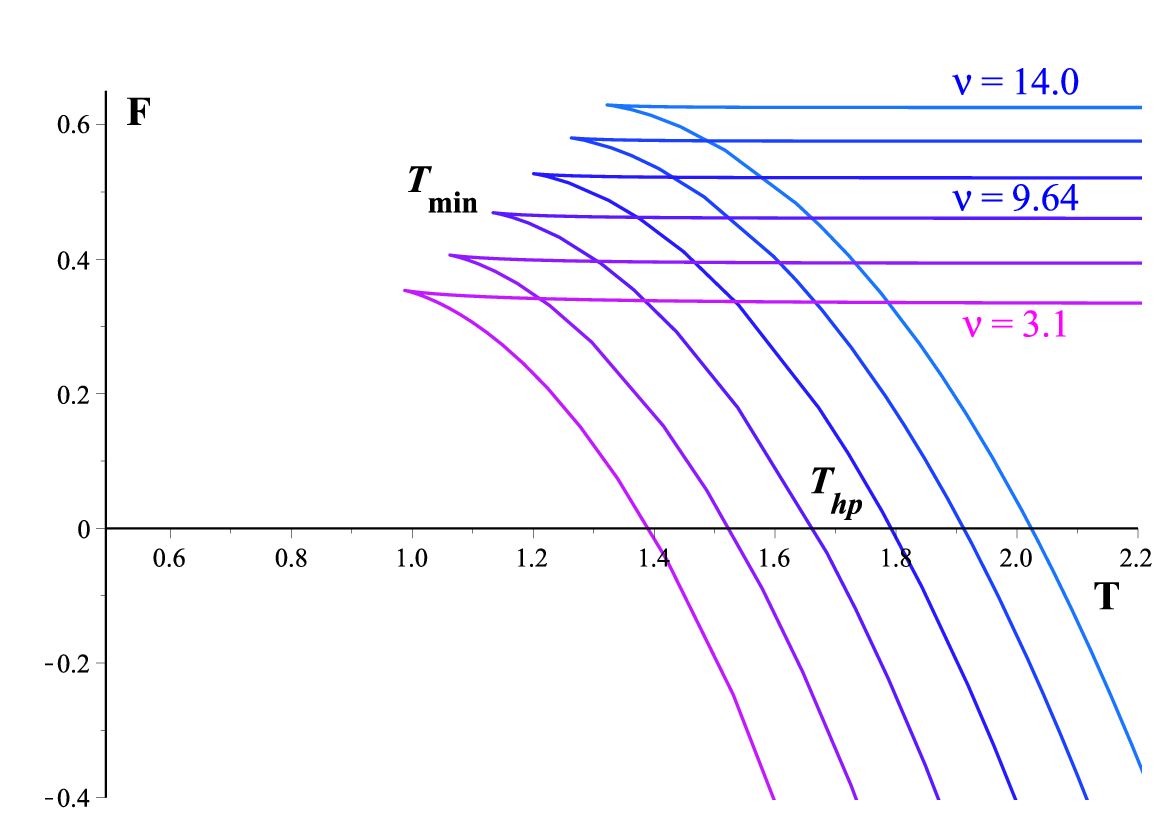}
}
\caption{The plots show the free energy as a function of temperature, illustrating the presence of a first-order phase transition at $F=0$. On the left, for $2.1\leq\nu\leq3$, the minimum temperature remains nearly constant. On the right, for $\nu>3$, the scalar hair significantly affects both the minimum temperature and the Hawking--Page temperature.
} 	
\label{fig2}  
\end{figure}

In Fig.~\ref{fig2} we again plot the free energy vs temperature relation. While the plots are similar with the ones of Schwarzschild--AdS black hole solution, the minimum temperature and Hawking--Page temperature are affected by the scalar hair as we can concretely observe for various values of the hair parameter $\nu$.   
On the left-hand side of Fig.~\ref{fig2}, we see that, for small amounts of scalar hair, the minimum temperature for which the hairy black holes exist does not vary significantly. However, on the right-hand side of Fig.~\ref{fig2}, we plot the case for which the hair parameter is $\nu > 3$ and observe that both temperatures grow directly proportional to the scalar hair. Intuitively, since there is a competition between the gravitational attraction and thermal fluctuations, the equilibrium settled at a higher temperature when there is more scalar hair.

\begin{figure}[!ht]
\centering
\subfloat{
\includegraphics[width=0.475\textwidth]{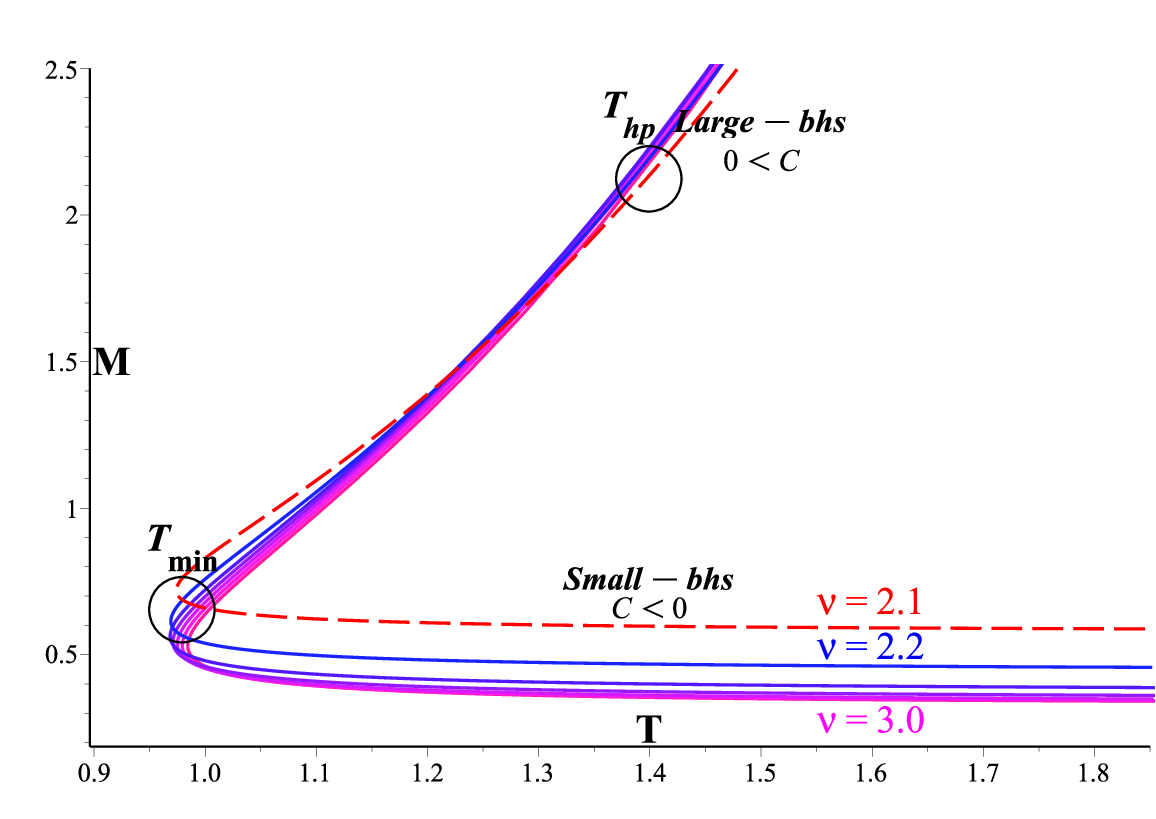}
}
\hfill
\subfloat{
\includegraphics[width=0.475\textwidth]{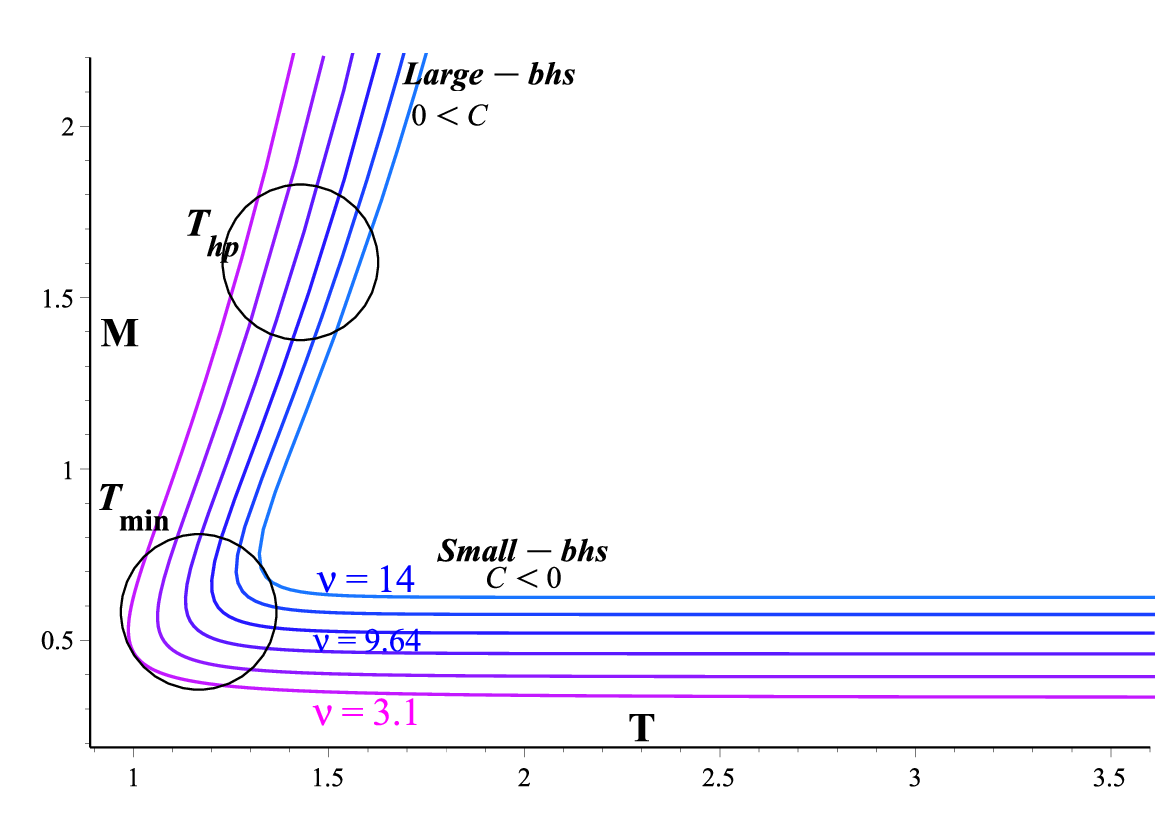}
}
\caption{The left panel ($2.1\leq\nu\leq 2.5$) shows the mass–temperature relation, where a discontinuity in the heat capacity at $T_\text{min}$ confirms the existence of two distinct branches, corresponding to small and large black holes. In the right panel ($\nu\geq3.1$) the temperature at which hairy black holes exist increases consistently with the hair parameter $\nu$.}    
\label{fig3}
\end{figure}

We complete the analysis with a plot of mass vs temperature in Fig.~\ref{fig3}, where the existence of the thermodynamically stable large black hole branch appears explicitly.

\section{Thermal superpotential and holographic interpretation}
\label{thermalsup}
The main goal of this section is to provide a holographic interpretation of the hairy black hole solutions in the context of AdS--CFT duality. The main results are obtaining the thermal superpotential in analytic form, using it as a counterterm to regularize the Euclidean action, and computing the dual stress tensor. As in \cite{Astefanesei:2021ryn}, where the $R$-charged black holes were analyzed in a similar way, there is no need for extra finite counterterms.

We start by generalizing the approach of \cite{Gursoy:2008za} on the thermal superpotential and first-order formalism to hairy black hole solutions with a spherical horizon topology. While the equations for the dilaton field and thermal superpotential remain unchanged as in the case of planar hairy neutral black hole solutions \cite{Anabalon:2020qux},
\begin{equation}
\frac{\varphi^{\prime}}{\eta\sqrt{\Upsilon}}=\frac{d\mathcal{W}}{d\varphi}, \qquad \mathcal{W}(\varphi)=-\frac{\Upsilon^{\prime}}{\eta\Upsilon^{3/2}}
\end{equation}
the equation for the dilaton potential depends on the horizon topology and reads
\begin{equation}
V= \frac{1}{2}\left[\left(\frac{d\mathcal{W}}{d\varphi}\right)^{\!2}-\frac{3}{4}\,\mathcal{W}^{2}\right]f
+\frac{\mathcal{W}f^{\prime}}{2\eta\sqrt{\Upsilon}}+\frac{k}{\Upsilon}
\end{equation}
where $k=1/L^2$ and $k=0$ represent the spherical and toroidal horizon topology, respectively.

For our case, the positive branch of Family 2 ($\varphi>0$ or $x>1$), we obtain
\begin{align}
\label{superpot}
\mathcal{W}(\varphi)&=
	\frac{1}{L\,\nu}\left((\nu+1)\, e^{\varphi\,\ell\,\left(\frac{\nu-1}{2}\right)}+
	(\nu-1)\, e^{-\varphi\,\ell\,\left(\frac{\nu+1}{2}\right)}\right),
\\[2ex]
\frac{d\mathcal{W}}{d\varphi}&=\frac{1}{L\,\nu\,\ell}\left(e^{\varphi\,\ell\,\left(\frac{\nu-1}{2}\right)}-e^{-\varphi\,\ell\,\left(\frac{\nu+1}{2}\right)}\right),
\end{align}
where, as a reminder, $L$ is the AdS radius.

Let us now use the thermal superpotential (\ref{superpot}) as a counterterm and compare the results with the ones in Section \ref{DilatonCounterterm} where we have used specific dilaton counterterms. While in Section \ref{DilatonCounterterm}, the gravitational counterterm, $I_\mathrm{g}^{\textsc{e}}$, contains two terms (\ref{Igrav}), the superpotential contains intrinsically the first term determined only by the determinant of
the induced metric and so, to regularize the Euclidean action, in this section we are going to use only the following counterterms:
\begin{equation}
I_\text{ct}^\textsc{e}=\frac{1}{8\pi G}\,\int_{\partial M} d^{3}x\,\sqrt{h^\textsc{e}}\,\left(
	\frac{\mathcal{W}(\varphi)}{2L}+\frac{L}{2}\,R(h)\right)\;=\frac{\sigma_{k}\,L^{2}}{8\pi GT}\left(\frac{2r^{3}}{L^{4}}+\frac{2r}{L^{2}}+\frac{1}{2}\varphi_{0}^{2}\,r-\frac{2}{3}L^{2}\varphi_{0}^{3}\,\ell-\mu L^{2}\right).
\end{equation}

Therefore, the regularized Euclidean action is 
\begin{equation}
I^{\textsc{e}}\left(g^{\textsc{e}},\varphi\right)=I^{\textsc{e}}_\text{bulk}+I^{\textsc{e}}_{\textsc{gh}}+I_\text{ct}^\textsc{e}\:.
\end{equation}

By expanding all the terms at the boundary and considering also the contributions from the horizon, we obtain 
\begin{equation}
I^{\textsc{e}}_\text{bulk}+I^{\textsc{e}}_{\textsc{gh}}=-\frac{\mathcal{A}}{4G}+\frac{\sigma_{k}\,L^{2}}{8\pi G\,T}\left(-\frac{2r^{3}}{L^{4}}-\frac{2r}{L^{2}}-\frac{1}{2}\varphi_{0}^{2}\,r+\frac{2}{3}L^{2}\varphi_{0}^{3}\,\ell+2\mu L^{2}\right)
\end{equation}
and
\begin{equation}
I_\text{ct}^\textsc{e} = \frac{\sigma_{k}\,L^{2}}{8\pi G\,T}\left(\frac{2r^{3}}{L^{4}}+\frac{2r}{L^{2}}+\frac{1}{2}\varphi_{0}^{2}r-\frac{2}{3}L^{2}\varphi_{0}^{3}\ell-\mu L^{2}\right)
\end{equation}
so that all the divergent terms cancel each other. Therefore, as expected, the regularized Euclidean action becomes
\begin{equation}
\label{action1}
I^{\textsc{e}}\,T=
\frac{ L^{4}}{2 G}\left(-\frac{1}{L^2 \eta}+\frac{\nu^{2}-4}{3\eta^{3}} \right) - \frac{T\mathcal{A}}{4G}
\end{equation}
identical to the quantum statistical relation previously obtained in (\ref{actionF}).

From the point of view of the field theory, using the boundary conditions (\ref{bconditions}), we observe that the black hole induces a triple trace deformation of the operator with conformal dimension $\Delta=1$:
\begin{equation}
\varphi_{1}=\frac{dw}{d\varphi_{0}} \quad\Longrightarrow\quad w(\varphi_{0})=-\frac{\ell\,\varphi_{0}^{3}}{6}=\frac{\varphi_{0}\,\varphi_{1}}{3}=-\frac{1}{6\,\ell^{2}\,\eta^{3}}\:.
\end{equation}

The deformation yields sensible infrared dynamics and this boundary condition makes the on-shell Euclidean action of a
probe field more negative, as it is shifted by the quantity
\begin{equation}
I_\textsc{cft}  \rightarrow I_\textsc{cft}   + \int \sqrt{\gamma}\,w(\varphi_0)\:.
\end{equation}

The components of the quasi-local stress tensor (\ref{BY}) are
\begin{equation}
\tau_{tt}=\frac{L}{r}~\frac{\mu L^{2}}{8\pi G}+O\left(r^{-2}\right)\,,
\qquad 
\tau_{\theta\theta}=\frac{L}{r}~\frac{\mu L^{2}}{16\pi G}+O\left(r^{-2}\right)\,,
\qquad 
\tau_{\phi\phi}=\tau_{\theta\theta}\,\sin^2\theta\,.
\end{equation}

The boundary metric and the metric of the geometry where the dual field theory lives are related up to a conformal factor as
\begin{equation}
\gamma^{\text{bound}}_{ab}dx^{a}dx^{b}=\frac{r^{2}}{L^{2}}\left(dt^{2}-d\theta^{2}-\sin^{2}{\theta}\,d\phi^{2}\right)=\frac{r^{2}}{L^{2}}\gamma^{\text{dual}}_{ab}\:.
\end{equation}

Then, the boundary Brown-York stress tensor (\ref{BY}) is related to the dual stress tensor as follows \cite{Myers:1999psa}:
\begin{equation}
\langle\tau_{ab}^{\text{dual}}\rangle=\lim_{R\rightarrow\infty}\frac{R}{L}\tau_{ab}
\end{equation}
and its components are
\begin{equation}
\langle\tau_{tt}^{\text{dual}}\rangle=\frac{\mu L^{2}}{8\pi G}\,, \qquad \langle\tau_{\theta\theta}^{\text{dual}}\rangle=\frac{\langle\tau^{\text{dual}}_{\phi\phi}\rangle}{\sin^{2}{\theta}}=\frac{\mu L^{2}}{16\pi G}\:.
\end{equation}

Consider now a boundary-comoving observer with the four-velocity
$u_{a}=\delta_{a}^{0}$. Then, the dual stress tensor $\langle\tau_{ab}^{\text{dual}}\rangle$ takes the form of a perfect fluid
\begin{equation}
\langle\tau_{ab}^{\text{dual}}\rangle=(\rho+p)\,u_{a}\,u_{b}-p\,\gamma_{ab}\:,   
\end{equation}
where the density energy and pressure are
\begin{equation}
p=\frac{\mu L^{2}}{16\pi G},  \qquad \rho=\frac{\mu L^{2}}{8\pi G}=\frac{L^{2}}{8\pi G}\left(-\frac{1}{\eta L^{2}}+\frac{\nu^{2}-4}{3\eta^{3}}\right).
\end{equation}

We obtain its final expression in a compact form as
\begin{equation}
\langle\tau_{ab}^{\text{dual}}\rangle=\frac{\mu L^{2}}{16\pi G}\left(3\,\delta_{a}^{0}\,\delta_{b}^{0}-\gamma^{\text{dual}}_{ab}\right)\,.
\end{equation}

This stress tensor is covariantly conserved and its trace vanishes, $\gamma^{ab}\langle\tau_{ab}^\text{dual}\rangle=0$, as expected from the fact that
the boundary conditions preserve the conformal symmetry.

\section{Conclusions}
\label{Sec5}
In this paper, we have studied exact neutral hairy black hole solutions with spherical horizon geometry in $\mathcal{N}=2$ supergravity with FI terms with non-trivial dilaton potential.%
\footnote{Similar solutions \cite{Acena:2012mr, Anabalon:2020qux}, but with planar horizon geometry, were used in the context of AdS--CFT duality to describe exact holographic RG flows.
} 
We were interested in the pure electric FI sector in which the dilaton potential can be obtained from a real superpotential and we can use the first-order formalism to explicitly obtain counterterms, which regularize the Euclidean action. On general grounds, the counterterms are not unique corresponding to different regularization schemes, but they should provide an invariant result of the thermodynamic potential. There is no need for finite counterterms and the asymptotic expansion is consistent with the results of \cite{Marolf:2006nd,Anabalon:2015xvl}.

\begin{figure}[!ht]
\centering
\subfloat{
\includegraphics[width=0.475\textwidth]{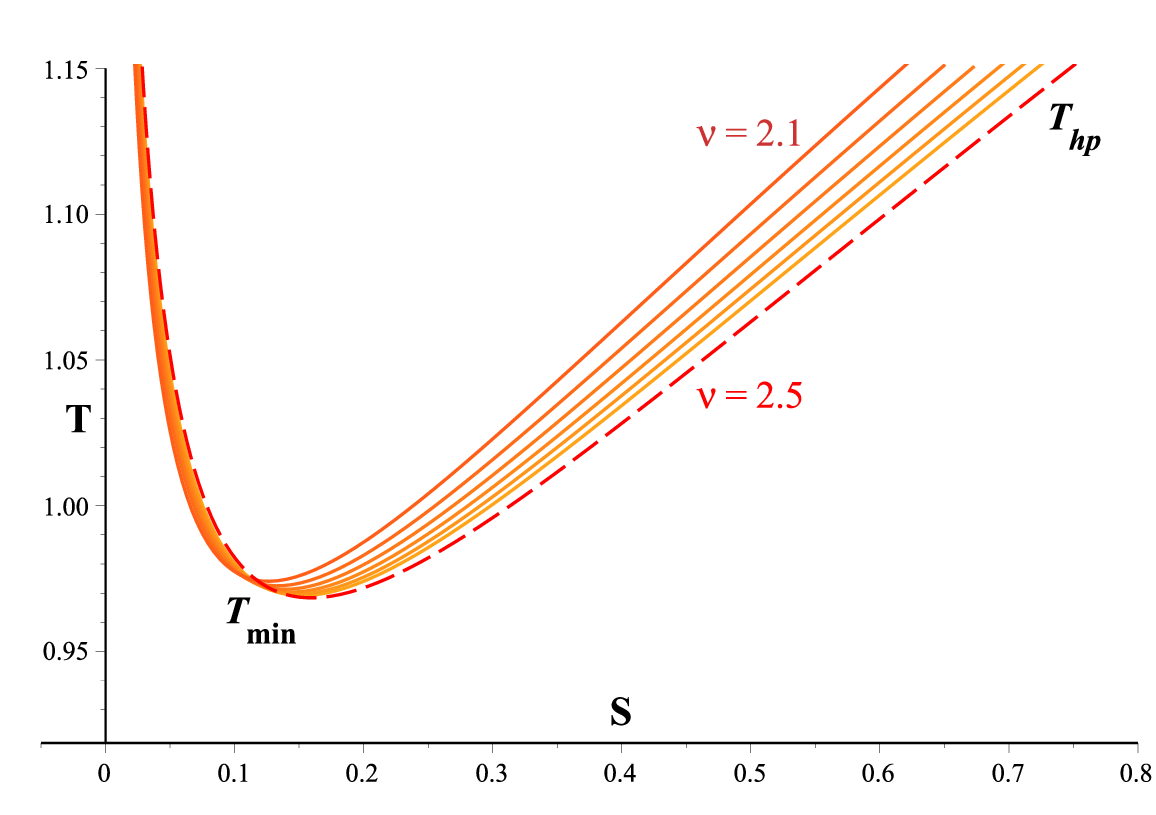}
}
\hfill
\subfloat{
\includegraphics[width=0.475\textwidth]
{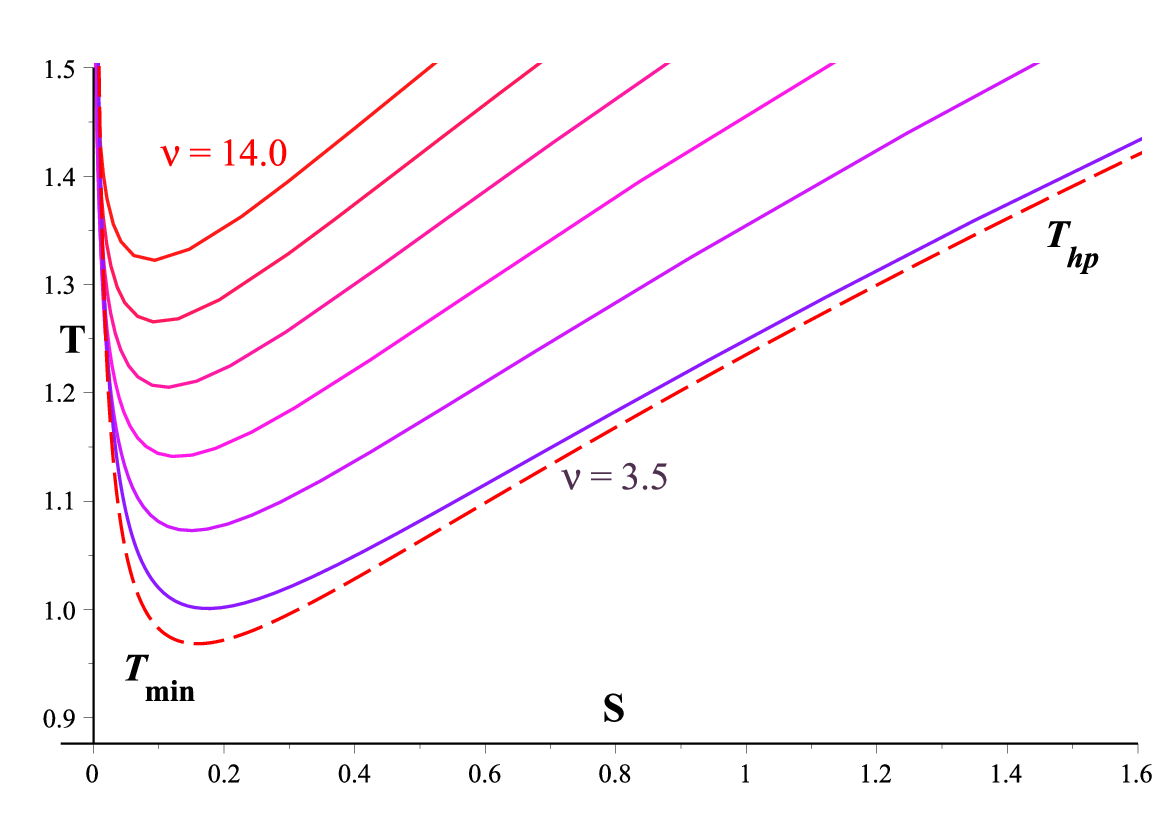}
}
\caption{The left panel ($2.1 \leq \nu \leq 2.5$) shows that the minimum temperature $T_\text{min}$ does not vary significantly. In contrast, in the right panel ($\nu \geq 3.5$) $T_\text{min}$ grows more rapidly as the scalar hair increases.}
\label{fig4}
\end{figure}

The thermodynamic behavior is similar to that of the Schwarzschild--AdS black hole \cite{Hawking:1982dh}, though the hair parameter in the dilaton potential, $\nu$, controls the thermodynamic quantities. We emphasize that the hair is secondary in the sense that there is no conserved charge associated with. In Fig.~\ref{fig4}, where 
we present the plot of the temperature vs entropy, we can identify a minimum temperature for which the black hole exists. For each temperature, above the minimum temperature, there exist two black holes, though just the large one is thermodynamically stable. We have proved that there exist phase transitions of first 
order between the stable black hole and thermal AdS. In Fig.~\ref{fig4}, we observe that the effect of the scalar hair becomes significant when the hair parameter is $\nu>3$.

\begin{figure}[!ht]
\centering
\subfloat{
\includegraphics[width=0.55\textwidth]{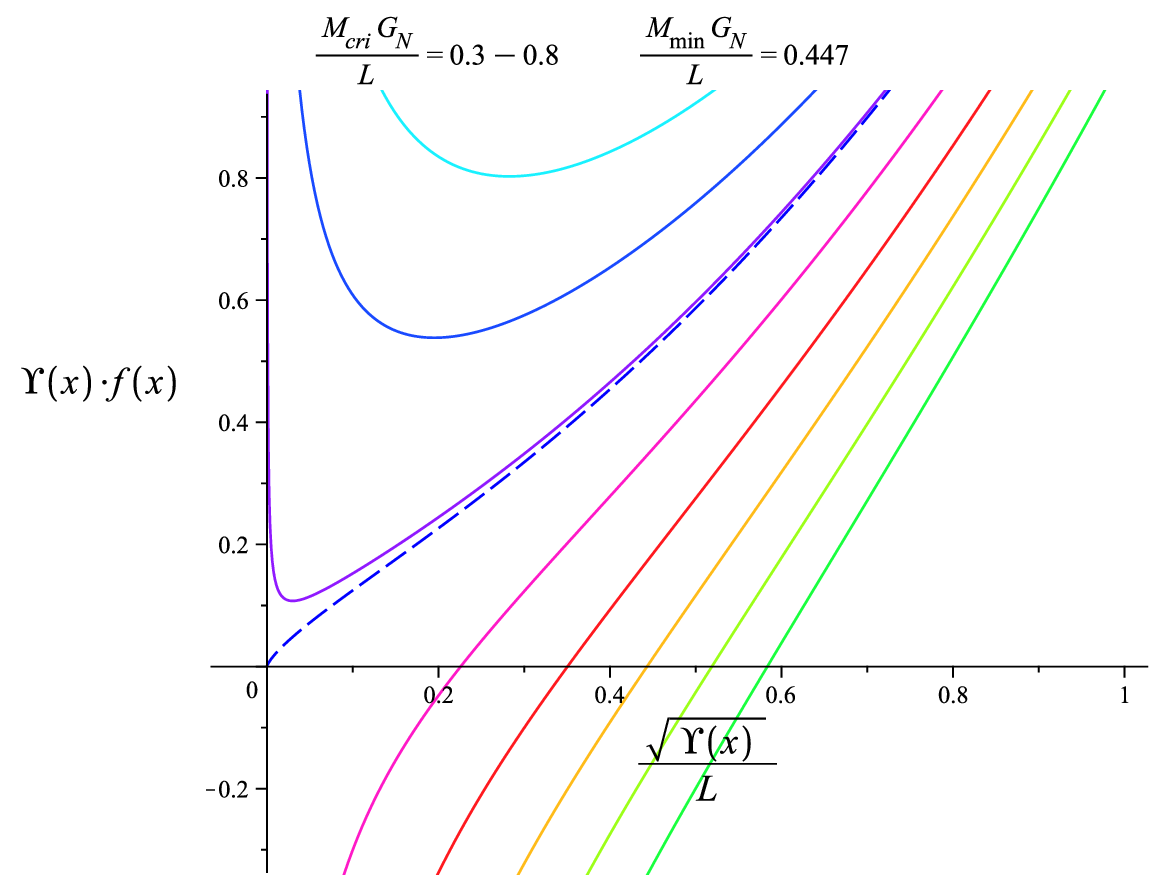}
}
\hfill
\subfloat{
\includegraphics[width=0.4\textwidth]{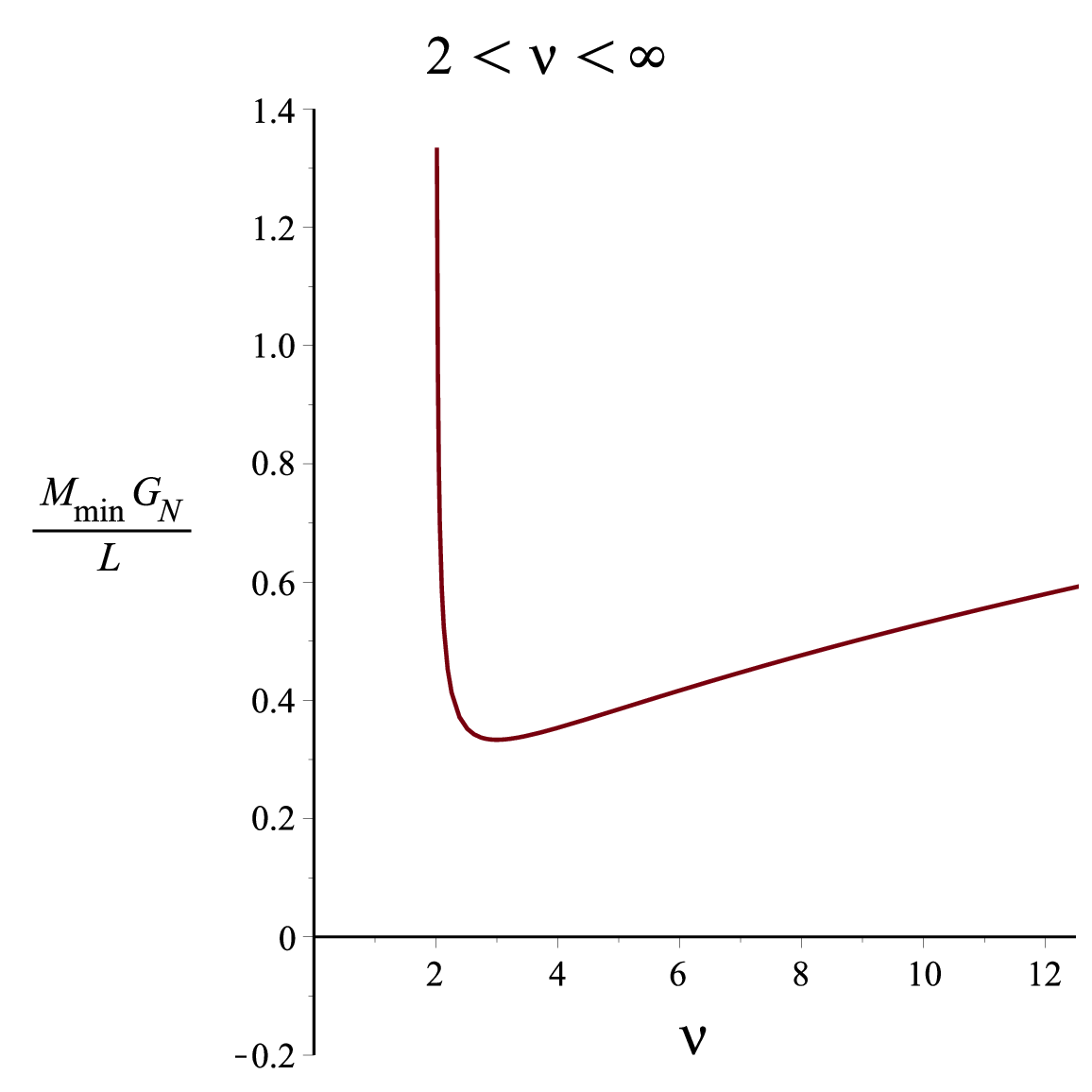}
}
\caption{In the left panel ($\nu=2.2$), the plots of $g_{tt}$ versus $\Upsilon$ for various values of $M$ reveal the presence of the event horizon. The dashed curve represents $M_\text{min}\,G/L=0.447$, indicating that an event horizon forms only when $M>M_\text{min}$. The right panel shows the dependence of $M_\text{min}$ on $\nu$, exhibiting a minimum at $\nu=3$ with $M_\text{min}=L/(3G)$.}
 \label{fig5}
\end{figure}

The embedding in supergravity makes the theory well defined with a stable ground state. 
Using $\eta_\text{cr}$ obtained in eq.\ (\ref{F2}), we get 
\begin{equation}
M\,>\, M_\text{min}(\eta_\text{cr})\,=\,
\frac{\sigma_{+1}\,L}{8\pi G}\left(\frac{\nu-1}{3\sqrt{\nu-2}}\right)=\,
\frac{L}{6G}\frac{\nu-1}{\sqrt{\nu-2}}\,\geq\,\frac{L}{3G}\:,
\label{F3}
\end{equation}
the bound being saturated at $\nu=3$, yielding $M_\text{min}=L/3G$. This is depicted in Fig.~\ref{fig5} where one can explicitly see that, for a fixed value of the hair parameter $\nu>2$, the $tt$-component of the metric vanishes only when $M>M_\text{min}$.

Under dilaton boundary conditions, $M_\text{min}$ is the branch-existence threshold -- i.e., the minimal mass of regular hairy black holes in this family -- and it effectively enforces cosmic censorship for configurations with $\nu>2$ (positive branch $x>1$).
With the mixed boundary conditions for the dilaton, the dual field theory is deformed by a triple-trace operator and the holographic stress tensor is that of a thermal conformal fluid in ultraviolet: in this ensemble, $M_\text{min}$ represents the onset of a deconfined hairy phase that admits a horizon. Due to the presence of the horizon, there is no infrared fixed point as in domain-wall case, and so the dual field theory has a mass gap.

As a future direction, our interpretation of hair/boundary deformations as turning on VEVs and selecting different IR completions also suggests exploring possible horizonless configurations (soliton limit) within our branch of solutions, along the lines of \cite{Anabalon:2022aig, Anabalon:2024qhf,Anabalon:2023oge}, and comparing the resulting energy with that of thermal AdS and black hole solutions. The latter models, in fact, share the same truncation origin and superpotential/first-order formulation, 
while featuring an analogous ensemble dependence of minimum energy, with the same focus on interpreting lower-energy bounds in AdS--CFT.

\section*{\normalsize Acknowledgments}
\vspace{-5pt}
A.A.\ is supported in part by the FONDECYT grants 1200986, 1210635, 1221504, 1230853 and 1242043. A.A.\ is an Alexander von Humboldt fellow. The work of D.A.\ is supported in part by FONDECYT grants 1242043, 1250133, 1211545, and 1240955.


\newpage

\hypersetup{linkcolor=blue}
\phantomsection 
\addtocontents{toc}{\protect\addvspace{4.5pt}}
\addcontentsline{toc}{section}{References} 
\bibliographystyle{JHEP}
\bibliography{bibliographyBH} 

\end{document}